\pdfoutput=1

\documentclass[11pt,a4paper,onecolumn]{article}

\usepackage{cite}
\usepackage{amsmath,amssymb,amsfonts}
\usepackage{algorithmic}
\usepackage{graphicx}
\usepackage{textcomp}
\usepackage{xcolor}
\usepackage{hyperref}

\usepackage[inline]{enumitem}
\usepackage{siunitx}
\usepackage{bm}
\renewcommand{\vec}{\bm}

\usepackage[left=30mm, right=30mm]{geometry}

\providecommand{\keywords}[1]{\textbf{\textit{Keywords---}} #1}

\begin{document}

\title{Uncertainty propagation in phaseless electric properties tomography}

\author{Alessandro Arduino \and Oriano Bottauscio \and Mario Chiampi \and Luca Zilberti}

\date{Metrologia dei materiali innovativi e scienze della vita\\
Istituto Nazionale di Ricerca Metrologica (INRiM)\\
Torino, Italy}

\maketitle

\noindent{\bf Acknowledgement:} The project 18HLT05 “Quantitative MR-based imaging of physical biomarkers” leading to this application has received funding from the EMPIR programme co-financed by the Participating States and from the European Unions Horizon 2020 research and innovation programme.
\\\\\\
\textsuperscript{\textcopyright} 2019 IEEE. This is the author’s version of an article that has been published by IEEE. Personal use of this material is permitted. Permission from IEEE must be obtained for all other uses, in any current or future media, including reprinting/republishing this material for advertising or promotional purposes, creating new collective works, for resale or redistribution to servers or lists, or reuse of any copyrighted component of this work in other works.\\
\\\\
Published in:\\
2019 International Conference on Electromagnetics in Advanced Applications (ICEAA)\\
Date of Conference: 9-13 Sept. 2019\\
Date Added to IEEE Xplore: 24 October 2019\\
Electronic ISBN: 978-1-7281-0563-5\\
USB ISBN: 978-1-7281-0562-8\\
Print on Demand(PoD) ISBN: 978-1-7281-0564-2\\
DOI: 10.1109/ICEAA.2019.8879147\\
Publisher: IEEE\\
Conference Location: Granada, Spain, Spain\\
\\\\
Available at: (DOI): \href{www.doi.org/10.1109/ICEAA.2019.8879147}{10.1109/ICEAA.2019.8879147}

\newpage

\maketitle

\begin{abstract}
Uncertainty propagation in a phaseless magnetic resonance-based electric properties tomography technique is investigated using the Monte Carlo method.
The studied inverse method, which recovers the electric properties distribution at radiofrequency inside a scatterer irradiated by the coils of a magnetic resonance imaging scanner, is based on the contrast source inversion technique adapted to process phaseless input data.
\end{abstract}


\section{Introduction}
The possibility to perform quantitative imaging of the electric properties (EPs) at radiofrequency (RF) of a scatterer (in particular, a human body) irradiated by the RF magnetic field generated and measured by a magnetic resonance imaging (MRI) scanner has been deeply investigated in the last years~\cite{liu2017}.
All the methods proposed in literature to this end are generally indicated as magnetic resonance-based EPs tomography (MR-EPT).

MR-EPT is attracting an ever-growing interest within both the scientific and the medical communities because of its high resolution with respect to similar techniques (e.g., microwave tomography)~\cite{rahimov2017}.
A fully developed MR-EPT would allow the adoption of non-invasive physical biomarkers for the detection and the characterisation of some pathologies, like breast cancer~\cite{liu2017}.
Moreover, it would pave the way to personalised medicine, allowing to perform patient-specific planning of electromagnetic (EM)-based treatment, like oncological hyperthermia~\cite{liu2017}.

The main issues to be overcome in order to make MR-EPT a clinically relevant medical tool are:
\begin{enumerate*}
    \item the impossibility to measure the phase of the input data without symmetry assumptions on the geometry of the used RF coil and of the imaged body~\cite{liu2017};
    \item the lack of information on the uncertainty associated to the recovered EPs maps (just few papers~\cite{lee2015,arduino2017} deal with that).
\end{enumerate*}

The first mentioned issue is particularly relevant when multi-channel RF coils for parallel transmission (pTx) MRI are considered. In general, a pTx coil does not fulfill the symmetry assumption required in order to estimate the input phase.
On the other hand, pTx coils allow the acquisition of multiple independent measurements, one for each transmit channel. Thus, different strategies for the implementation of MR-EPT with pTx MRI have been proposed that overcome the phase measurement issue by recovering the lacking phase information thanks to the multiple available inputs~\cite{arduino2018,bevacqua2019,serralles2019}.
Amongst these methods, the contrast source inversion (CSI) global Maxwell tomography (GMT)~\cite{arduino2018} takes advantage of a phaseless implementation of the CSI technique, originally developed for scattering inverse problems~\cite{vandenberg2001}.

In this paper, in order to obtain information about how the uncertainty propagates through CSI-GMT, the Monte Carlo method (MCM) is applied, in accordance to the Supplement 1 to the Guide to the Expression of Uncertainty in Measurement (GUM)~\cite{gumsupp1}, because of the technique's non-linearity and complexity.
In particular, a realistic two-dimensional virtual problem that models a human head irradiated by a transverse EM (TEM) RF coil for MRI with 4, 8 and 16 channels is used as the reference.
The quantified uncertainty in the recovered maps will be related to the number of available input measurements as well as to the input uncertainty, described in terms of signal to noise ratio (SNR).

\section{Method}
\begin{figure}[t]
\centerline{\includegraphics{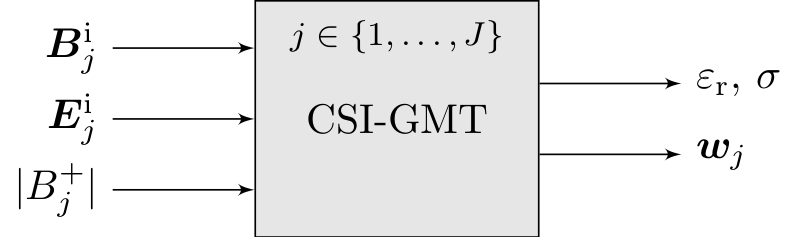}}
\caption{Diagrammatic representation of the CSI-GMT method.}
\label{fig:CSI-GMT}
\end{figure}

CSI-GMT is a global MR-EPT method that recovers the spatial distribution of the EPs within the examined domain by minimising a proper cost functional. Implementative details of the iterative procedure could be found in~\cite{arduino2018}, but, for the analysis presented here, CSI-GMT could be just interpreted as the black-box function depicted in Fig.~\ref{fig:CSI-GMT}.

In Fig.~\ref{fig:CSI-GMT}, each channel of the pTx coil employed to provide the input values is denoted by the symbol $j = 1,\dots,J$.
CSI-GMT inputs are the incident EM field $\{\vec{E}^{\rm i}_j, \vec{B}^{\rm i}_j\}$ and the transmit sensitivity magnitude $|B_j^+|$ of each channel.
The incident EM field is the one that is generated by the $j$-th channel without the scatterer; whilst the transmit sensitivity is a rotating component of the RF field generated in presence of the scatterer, $B^+ = \left(B_x + {\rm i} B_y\right)/2$~\cite{hoult2000}. No use of $B^+$ phase is done by CSI-GMT~\cite{arduino2018}.

Uncertainty propagation through the CSI-GMT method is studied assuming that the incident EM field is perfectly known and only the transmit sensitivity magnitude is affected by random errors.
Indeed, $\{\vec{E}^{\rm i}_j, \vec{B}^{\rm i}_j\}$ can be evaluated off-line, for instance, by means of a highly accurate numerical simulation, with an expected standard uncertainty significantly lower than the one with which the {\it in vivo} measurements of $|B_j^+|$ could be acquired.
Possible misalignments between the input maps, which would invalidate this assumption, are neglected.

Because of the non-linearity and the complexity of CSI-GMT, the uncertainty in the output EPs, $\varepsilon_{\rm r}$ and $\sigma$ (the relative permittivity and the electric conductivity, respectively), is evaluated using the Monte Carlo method (MCM)~\cite{gumsupp1}.
The uncertainty in the auxiliary unknown $\vec{w}_j$, which provides information about the RF EM field generated inside the scatterer, is not analysed in this paper.
$M = 1000$ sample draws are used when performing the MCM. The number of used samples is smaller than the one prescribed by the Supplement 1 to GUM~\cite{gumsupp1}, but it is forced by the computational cost of the CSI-GMT implementation. Anyway, looking at the approximated probability density functions resulting from the MCM application, it can be stated {\it a posteriori} that 1000 samples are enough to satisfactorily evaluate the uncertainty propagation in the considered cases.

Virtual measurements of the input transmit sensitivity magnitude are modelled as
\begin{equation}\label{eq:virtual_measurements}
    |B_j^+|_{\rm meas} = \left| |B_j^+| + n \right|\,,
\end{equation}
where $|B_j^+|$ is the actual transmit sensitivity magnitude of the $j$-th channel of the RF coil and $n(\vec{x}) \overset{\rm iid}{\sim} \mathcal{N}(0,u^2)$, $\forall \vec{x}$ is the random error at the acquisition points.
The random error is assumed spatially uncorrelated, since, to the best of the Authors' knowledge, nothing about its spatial correlation has been documented, yet.
The standard deviation $u$ is choosen proportional to the spatial average of $|B_j^+|$ according to the relation $u = {\rm mean}(|B_j^+|)/{\rm SNR}$, where SNR is the signal-to-noise ratio of the sensitivity map.
The absolute value in~(\ref{eq:virtual_measurements}) na\"ively models any generic B1-mapping technique for the transmit sensitivity magnitude acquisition.

A realistic two-dimensional model problem is considered as reference for the MCM. In the model problem, a section of the head of the anatomical human model 'Duke' from the Virtual Population~\cite{gosselin2014} is irradiated by TEM coils for pTx MRI with $J = 4$, $8$ and $16$ channels.
The RF field generated by the coil at \SI{128}{MHz} (the operative frequency of \SI{3}{T} MRI scanners) with and without the head section is simulated numerically, modelling the coil channels as couple of line sources and using the method of moments, as already done elsewhere~\cite{arduino2018}. The obtained results are used as the exact values and are properly corrupted to perform the Monte Carlo analysis.
The EPs at \SI{128}{MHz} assigned to the head tissues are taken from the IT'IS foundation database~\cite{hasgall2018}.

\section{Results and discussion}
\begin{figure}[t]
\centerline{\includegraphics{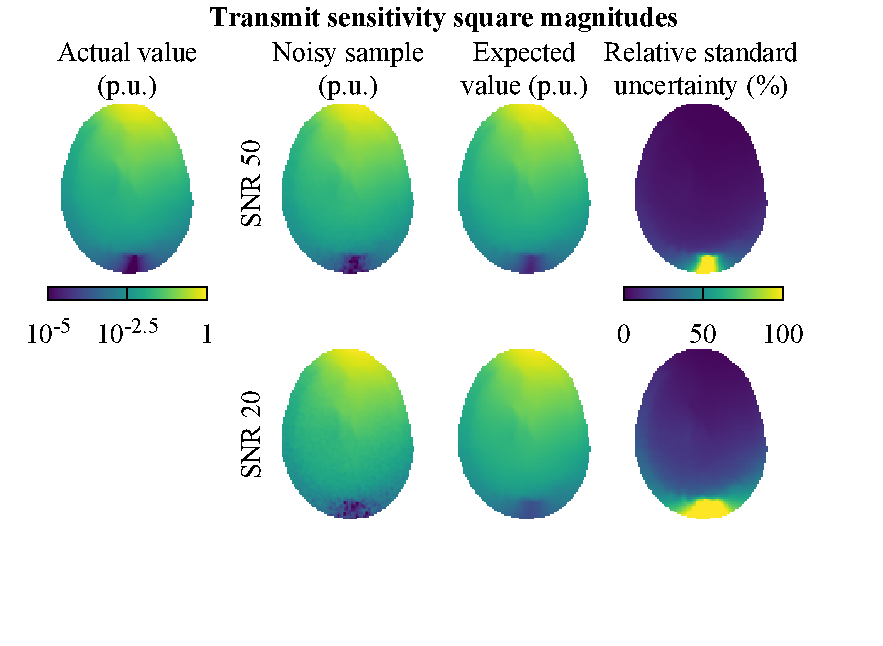}}
\vspace{-2.3cm}
\hspace{3.5cm}\includegraphics[width=2cm]{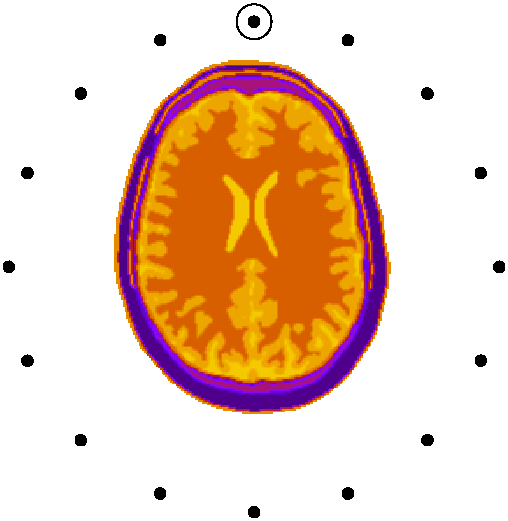}
\caption{From left to right, map of the actual transmit sensitivity square magnitude, examples of corrupted input with SNR 20 and 50, expected values and relative standard uncertainty computed using the MCM. The maps refers to the transmit sensitivity associated to the channel located in front of the head, as indicated by the inset on the bottom left where a 16-channels coil is depicted.}
\label{fig:input}
\end{figure}

The corrupted input for the Monte Carlo procedure are illustrated in Fig.~\ref{fig:input}, where the actual transmit sensitivity square magnitude of a channel located in front of the head is reported together with noisy samples and spatially distributed statistics.
It is worth noting that the virtual measurement $|B_j^+|_{\rm meas}$, defined as in~(\ref{eq:virtual_measurements}), is affected by a SNR-dependent bias. This fact is put in evidence from the expected values depicted in Fig.~\ref{fig:input}, which, in the nape, report a transmit sensitivity magnitude higher than the actual one. This is a consequence of the absolute value in~(\ref{eq:virtual_measurements}).
Moreover, the relative standard uncertainty is very low near the transmit source, where the maximum of the transmit sensitivity magnitude is located, and increases moving away from the source until the peak is reached (in the nape, for the considered channel).

\begin{figure}[t]
\centerline{\includegraphics{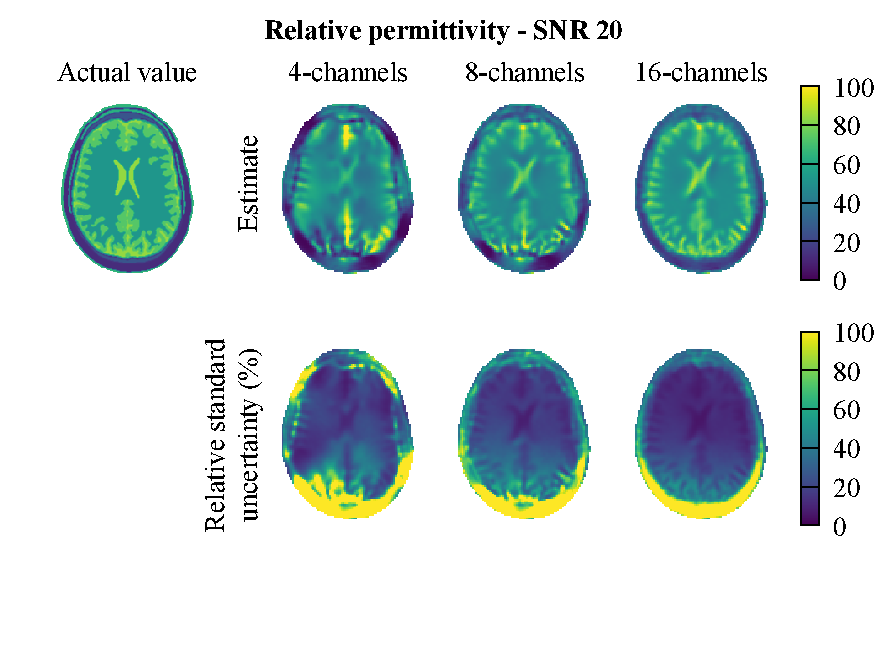}}
\vspace{-.5cm}
\centerline{\includegraphics{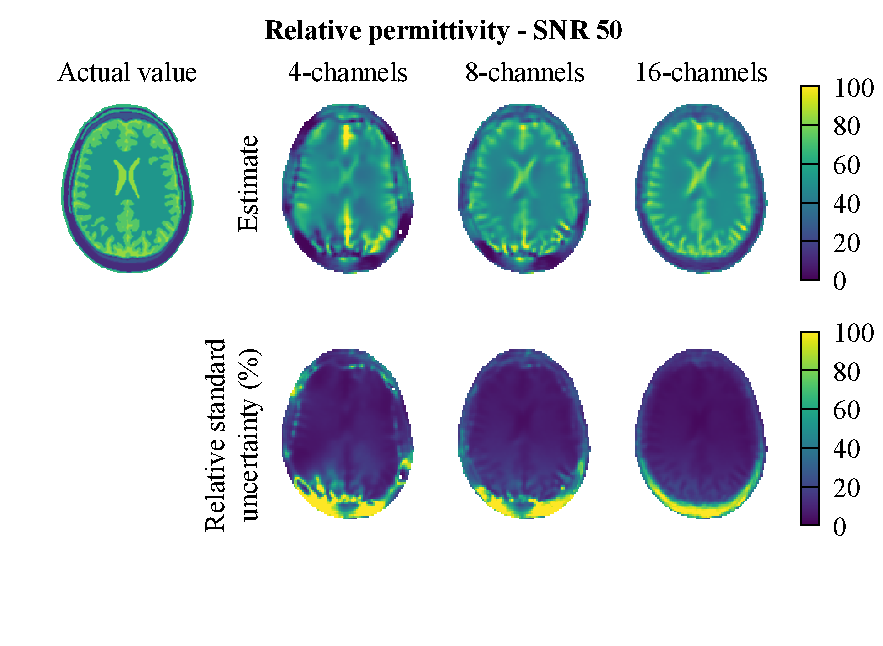}}
\vspace{-.8cm}
\caption{Maps of the actual and estimated distributions of the relative permittivity $\varepsilon_{\rm r}$ in the head section as well as of the spatial distributions of the relative standard uncertainty associated to the estimates. Results are reported when $J=4$, $8$ and $16$ channels are used to provide inputs with both SNR 20 and 50. Computations are performed according to the MCM with $M=1000$ extractions.}
\label{fig:epsr}
\end{figure}

\begin{figure}[t]
\centerline{\includegraphics{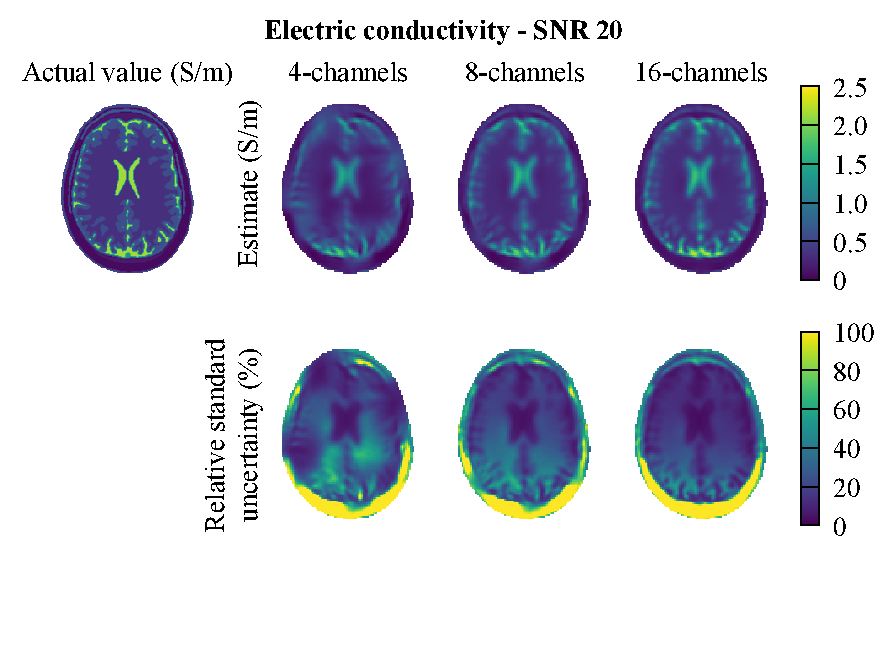}}
\vspace{-.5cm}
\centerline{\includegraphics{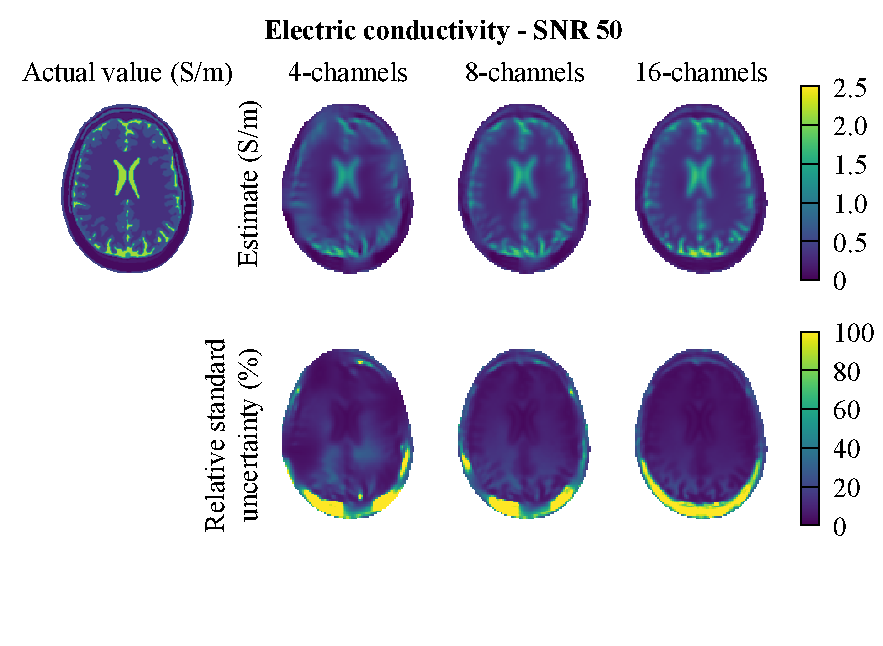}}
\vspace{-.8cm}
\caption{Maps of the actual and estimated distributions of the electric conductivity $\sigma$ in the head section as well as of the spatial distributions of the relative standard uncertainty associated to the estimates. Results are reported when $J=4$, $8$ and $16$ channels are used to provide inputs with both SNR 20 and 50. Computations are performed according to the MCM with $M=1000$ extractions.}
\label{fig:sigma}
\end{figure}

In Figs.~\ref{fig:epsr} and~\ref{fig:sigma}, the actual EPs (relative permittivity $\varepsilon_{\rm r}$ and electric conductivity $\sigma$, respectively) distributions within the head section are reported and compared with their estimates, that are obtained by averaging arithmetically the EPs maps recovered by CSI-GMT applied to the noisy inputs of the MCM~\cite{gumsupp1}.
For both the EPs, the estimated map improves sensibly when the number of input channels is increased. On the other hand, the error made on the estimate is only slightly affected by the input SNR. This fact suggests that the SNR-dependent bias in the input, observed in Fig.~\ref{fig:input}, does not affect the inverse procedure significantly.
This statement could not hold for lower values of the input SNR.

Differently from the estimates, the relative standard uncertainty, computed by the MCM as the ratio between the standard deviation of the output distribution and the estimate, decreases when less noise is present in the provided inputs as well as when more input channels are available.
Moreover, the relative standard uncertainty of the output is higher in the back than in the front of the head section. This is probably due to the positioning of the head within the TEM coil, which can be seen in Fig.~\ref{fig:input}. The distance of the nape from its nearest source is such that, in general, the input signals are lower there than in the rest of the head, making the EPs recovery in that region more critical.

A comparison between Fig.~\ref{fig:epsr} and Fig.~\ref{fig:sigma} shows that a higher bias is present in the electric conductivity estimate than in the relative permittivity one. In particular, the most relevant error committed in $\sigma$ recovery is about the cerebrospinal fluid (CSF) property. CSF has a very high electric conductivity with respect to the surrounding tissues, making its evaluation difficult for CSI-GMT, which requires many iterative steps in order to reach a reasonable approximation.
Since, as a regularisation technique for handling noisy inputs, CSI-GMT has been stopped after just 200 iterative steps for all the studied cases, CSF electric conductivity is strongly underestimated.

\begin{figure}[t]
\centerline{\includegraphics{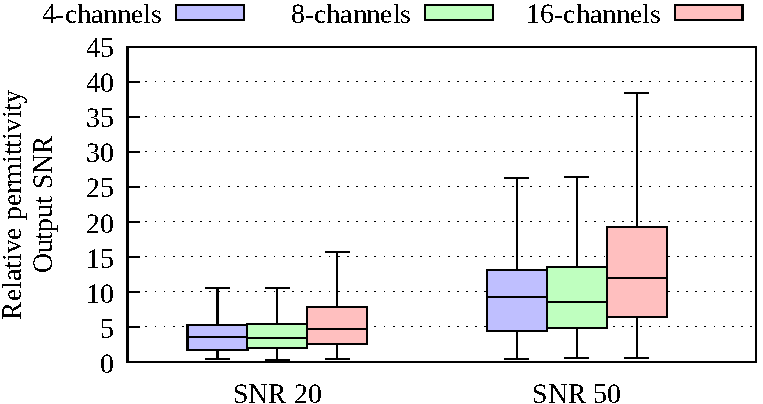}}
\vspace{.2cm}
\centerline{\includegraphics{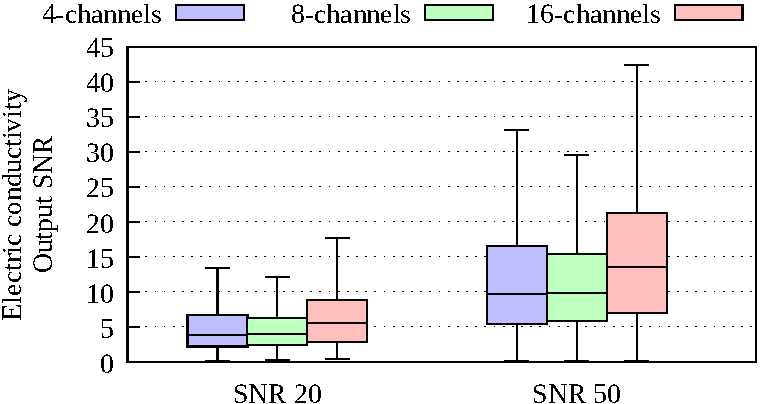}}
\caption{Boxplots summarizing the spatial information of the output SNR in the recovered EPs according to the MCM for different input SNR and channels. Output SNR is defined as the inverse of the relative standard uncertainty.}
\label{fig:boxplots_SNR}
\end{figure}

\begin{figure}[t]
\centerline{\includegraphics{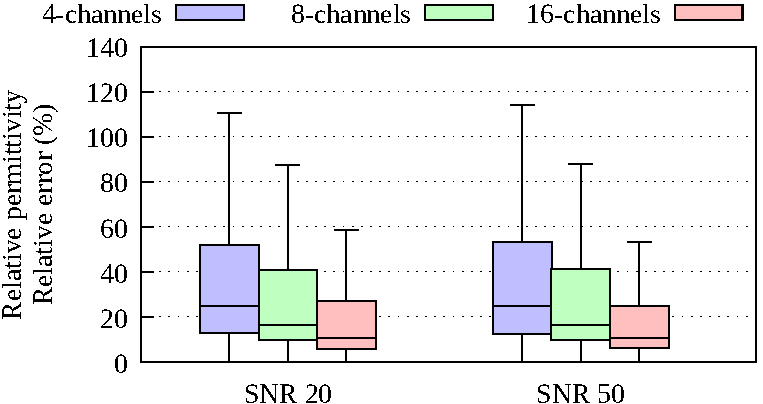}}
\vspace{.2cm}
\centerline{\includegraphics{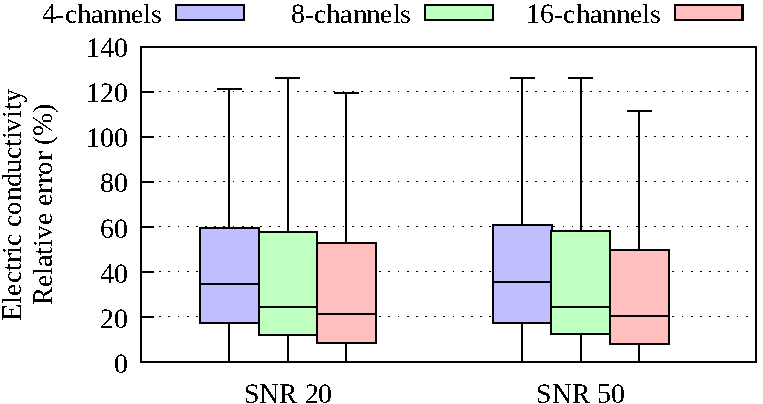}}
\caption{Boxplots summarizing the spatial information of the local relative error in the estimates of the recovered EPs according to the MCM for different input SNR and channels.}
\label{fig:boxplots_RelErr}
\end{figure}

In order to have a clearer and more synthetic way to interpret the spatially distributed statistics resulting from the MCM, they are collected in the boxplots depicted in Figs.~\ref{fig:boxplots_SNR} and~\ref{fig:boxplots_RelErr}.

In Fig.~\ref{fig:boxplots_SNR}, the spatial distribution of the output SNR, defined as the inverse of the relative standard uncertainty, is summarised in boxplots for each combination of input SNR and number of channels.
Despite moving from 8 to 16 input channels increases globally the output SNR for both the EPs, both 4 and 8-channels RF coils lead to almost the same SNR values in all the considered cases. This fact suggests a non-linear dependence of the output SNR with respect to the number of input channels.
On the other hand, a clear dependence of the output SNR to the input SNR can be appreciated.
In addition, almost the same output SNR is evaluated for both the EPs.

Finally, the local relative errors, computed as the relative difference between the estimate of each EP and its actual value, are summarised in Fig.~\ref{fig:boxplots_RelErr}, whose boxplots confirm that the bias in the recovered maps of both the EPs is almost independent of the input SNR.
On the other hand, from the same boxplots, the output bias appears to be proportional to the number of available transmit channels. This proportionality arises for both the EPs, despite the higher relative errors in the recovery of the electric conductivity, affected by the difficulty in estimating the very large conductivity of CSF.

\section{Conclusion}
In this paper, the MCM has been applied, in accordance to the Supplement 1 to GUM~\cite{gumsupp1}, in order to quantify the uncertainty propagation through CSI-GMT~\cite{arduino2018}, a global MR-EPT method based on the iterative minimisation of a cost functional that measures the discrepancy between the measured EM field and the one that would be generated in the presence of a guess distribution of EPs.
The main peculiarity of CSI-GMT with respect to other MR-EPT techniques is that it makes no assumptions on the phase of the input transmit sensitivity, by using only its magnitude~\cite{arduino2018}.

The analysis, performed with $M = 1000$ Monte Carlo extractions and stopping the iterative CSI-GMT procedure after 200 iterative steps as regularisation strategy, shows the dependence of the result accuracy to the input SNR and the number of available input channels.
In particular, a bias in the estimates of the EPs due to the early stop of the iterative procedure appears to be independent of the input SNR, despite an SNR-dependent bias in the inputs. The bias in the estimates decreases when the number of available input channels increases.

On the other hand, a higher input SNR leads to a less uncertain estimates of the EPs.
The uncertainty associated to the recovered properties depends on the number of input channels in a non-linear way. A sensible improvement is experienced only when moving from 8 to 16-channels transmit coils.

\end{document}